\documentclass{egpubl}
\usepackage{eurovis2026}

\EuroVisShort  

\usepackage[T1]{fontenc}
\usepackage{dfadobe}

\usepackage{cite}  

\usepackage{tabu}                      
\usepackage{booktabs}                  
\usepackage{tabularx}
\usepackage{forest}
\usepackage{amsmath}
\usepackage{makecell}
\usepackage{array}
\usepackage{mathptmx}                  
\usepackage{xcolor}
\usepackage[normalem]{ulem}

\BibtexOrBiblatex
\electronicVersion
\PrintedOrElectronic

\usepackage{egweblnk}

\title[Beyond Advocacy: A Design Space for Replication-Related Studies]%
      {Beyond Advocacy: A Design Space for Replication-Related Studies}

\author[Y. Liang, K. Marriott and H. C. Purchase]
{\parbox{\textwidth}{\centering Yiheng Liang\orcid{0009-0006-8029-2921}
        , Kim Marriott
        and Helen C. Purchase
        }
        \\
{\parbox{\textwidth}{\centering Department of Human-Centred Computing, Monash University, Australia
       }
}
}

\begin{document}


\maketitle
\begin{abstract}
   The importance of replication is often discussed and advocated -- not only in the domains of Visualization and HCI, but in all scientific areas. When replicating a study, design decisions need to be made regarding which aspects of the original study will remain the same and which will be altered (and how). We present a multi-dimensional design space framework (REPVIS) within which such decisions can be identified, categorized and analyzed. The framework treats replication experimental design as a pairwise comparison problem, and represents the design by four practical dimensions defined by three comparison levels. Our discussion of replication is therefore not simply advocacy: by presenting a design space we provide a concrete means by which all aspects of a replication study can be carefully planned, and which can be used to characterize and analyze existing studies.


\end{abstract}  
\section{Introduction}
Reproducibility and replicability have long been central concerns in scientific research~\cite{peng2011reproducible}. Replicating a previous study ensures the findings are validated, and can be cited with confidence~\cite{jasny2011again}. Replication in the physical sciences (e.g. the reaction when mixing two chemicals) is straightforward; not so in sciences where humans are part of the experiment. For example, both psychology and sociology suffer from insufficient replication to validate published findings~\cite{open2015estimating, freese_replication_2017}. So too for the fields of Visualization and HCI~\cite{kosara2018skipping, hornbaek2014once}.

The concerns that most published studies are not replicated, and that researchers are typically more interested in addressing new research questions than revisiting old ones are often discussed under the broader discourse of the "replication crisis"~\cite{kosara2018skipping,peng2015reproducibility}. Such reflection within the Visualization and HCI communities~\cite{hornbaek2014once,fekete2020exploring,quadri2019you} advocates for more replication~\cite{vis2024_opc_replication_2024}.

However, existing terminology and advocacy works alone provide limited support for the practical-level decisions that researchers must make when designing and conducting replication-related studies, and offer no concrete and consistent way to report how a new study (the replication study) relates to an existing one (the reference study). Our guidelines for designing single and independent experiments give little support for the process of designing a replication study, which naturally must correspond (in some way) to the reference study. Meaningful comparison requires making explicit how the replication study differs from the reference study, and what equivalence is (and is not) assumed. 

To facilitate this design requirement we propose a design space framework for replication studies (called REPVIS), making clear the details and design choices that researchers need to consider. Our work contrasts with existing research which takes a more abstract and less actionable approach to defining the scope of replication. Our REPVIS design space presents design choices as dimension-wise correspondences between a replication study and a reference study. Our goal is to show the full scope of replication study design possibilities, not to provide specific design recommendations (which might vary according to context). Our contributions are:
\begin{itemize}
    \item Reframing replication study design as comparison and correspondence between a replication study and a reference study;
    \item Describing the correspondence as a multi-dimensional space supporting planning and retrospective characterization;
    \item Illustrating use of the framework with worked examples.
\end{itemize}

Despite extensive discussion in the scientific fields~\cite{rougier_rescience_rwords_issue_2016,barba2018terminologies,acm_artifact_review_badging_2020}, terminology related to replication remains inconsistent across the Visualization and HCI domains. Terms such as \emph{replication} and \emph{reproduction} are often used with overlapping or conflicting meanings. Rather than attempting to resolve or enforce a particular terminology, we use \emph{replication} as an umbrella term to refer to studies that explicitly attempt to address the same (or similar) research question as a prior study, by conducting a new one. Our use of \emph{replication} is pragmatic rather than normative. It serves only to delineate our scope, not to prescribe a particular definition or to adjudicate between competing conventions.


\section{Related Work}
\subsection{Replication Discourse}
Concerns about the lack of reproducibility and replicability have been widely discussed across scientific disciplines, often framed through the broader "replication crisis"~\cite{open2015estimating,leek2015glass,cockburn2020threats}. While attempts have been made to define taxonomies and clarify terminology, naming conventions across communities are still incompatible and ambiguous~\cite{plesser2018reproducibility,drummond2009replicability}, and have been shown to have shifted over time and fields~\cite{barba2018terminologies}. The Visualization and HCI research communities have similar challenges~\cite{kosara2018skipping}, with discussion of replication highlighted in community venues such as the BELIV 2018 workshop~\cite{sedlmair2018proceedings} and the CHI 2013 RepliCHI workshop~\cite{wilson2013replichi}, alongside other reflective articles~\cite{sukumar2018towards,kosara2018skipping,lucke2018lowering,valdez2018requirements}. In addition, recent research still highlights the inconsistent usage~\cite{isenberg2024state}.


Beyond advocacy articles, publication of replicated experiments is, while still rare, increasing in the visualization literature~\cite{kay2015beyond,creamervalidation}. For example, multiple studies~\cite{davis2022risks,haehn2018evaluating} have replicated classic graphical perception results~\cite{cleveland1984graphical}, and the rise of crowdsourcing has enabled prior in-person experiments to be replicated online~\cite{heer2010crowdsourcing}. In addition, it is increasingly common that a study is replicated with participants taken from a different population to that used in the reference study~\cite{khalaila2025they,haehn2018evaluating}, thus testing the generalisability of the prior results. However, the reporting of these studies lacks a consistent and unambiguous framework for articulating the correspondence between the replication study and the reference study.

\subsection{Empirical Study Design}
Replication design is, fundamentally, an experimental design problem. Existing guidelines for the design of experiments in Visualization and HCI~\cite{purchase2012experimental,mackenzie2024human} assume a single, independent study. In contrast, replication study design is reference-constrained: it requires establishing pair-wise comparability and correspondence between the new study and a reference study at the level of practical choices.

Recent works which provide technical support for replication study implementation~\cite{ding2023revisit,cutler2026revisit}, suggest an educational method for conducting replication in class~\cite{DBLP:conf/vissym/SyedaSRPSMDB24}, and give some practical heuristics in the context of Visualization research ~\cite{sukumar2018towards} are broad and do not address practical-level decisions. With reference to biological research, Patil et al. propose a framework that operationalizes replication (same vs.\ different) across experimental design dimensions~\cite{patil2016statistical}. Overall, there is no concrete, well-defined framework that supports consistent characterization and planning for HCI and Visualization replication studies.

\section{A Replication Design Space Framework}
\label{sec:framework}

\subsection{Rationale}
Replication requires systematic comparability between the replication study and the reference study. While existing articles that emphasize advocacy, terminology discourse, or high-level taxonomies provide valuable context, they offer limited support for practical-level decisions and for reporting what \emph{changed}, what \emph{stayed comparable}, what is \emph{deliberately different}, and, ultimately, why the comparison remains meaningful. We therefore treat replication-related design as a reference-constrained, pairwise design problem.

Our design space (i) represents replication design through four study design dimensions and three comparison levels, and (ii) instantiates the result as an unambiguous multi-dimensional space that supports both retrospective characterization and prospective planning. The framework is general enough to cover both human studies and computation-only ones (\S\ref{sec:collapse_rules}), but is motivated by the complex needs of the former where practical guidance is scarce.




\begin{table*}[htbp]
\centering
\scriptsize
\renewcommand{\arraystretch}{1.25}
\setlength{\tabcolsep}{4pt}

\begin{tabularx}{\textwidth}{@{}
>{\raggedright\arraybackslash}p{0.12\textwidth}
*{4}{>{\raggedright\arraybackslash}X}
@{}}
\toprule
 & \colorbox[HTML]{B3E2CD}{Experiment} & \colorbox[HTML]{FDCDAC}{Data} & \colorbox[HTML]{CBD5E8}{Participant} & \colorbox[HTML]{F4CAE4}{Analysis} \\
\midrule

\textbf{\textit{Identical}} &
\textbf{Criterion:} Same experimental procedure, same stimuli, same task

\textbf{Example:} No change in experiment set up; unavoidable implementation differences are allowed&
\textbf{Criterion:} Same data form and data values

\textbf{Example:} Exactly the same data values provided by the reference study is used &
\textbf{Criterion:} Same population and same sampling method

\textbf{Example:} Participants have the same expertise and demographics, and the recruitment method is the same &
\textbf{Criterion:} Same approach of data analysis

\textbf{Example:} The same statistical methods are used

\\
\textbf{\textit{Similar}} &
\textbf{Criterion:} Some aspects of the experiment procedure are changed

\textbf{Example:} A different stimuli set is used, or different tasks are specified, or a between-participants design is used instead of a within-participants one &
\textbf{Criterion:} Data form is the same, but the data values are different

\textbf{Example:} Both the reference study and the replication study collect task accuracy and response time &
\textbf{Criterion:} Core, study-relevant population characteristics are aligned; non-relevant attributes may differ

\textbf{Example:} Only the age-range of the population is changed which is irrelevant to the skill requirement of study &
\textbf{Criterion:} The data is treated differently, but with a similar approach

\textbf{Example:} A multiple regression is used to determine factor effects, rather than ANOVA to investigate differences between them \\

\textbf{\textit{Different}} &
\textbf{Criterion:} The experimental paradigm is significantly different

\textbf{Example:} Research questions addressed by a lab study are investigated with a new study conducted in-the-wild &
\textbf{Criterion:} The new data collected is in a different form to the reference study data

\textbf{Example:} Qualitative data is collected rather than quantitative data. &
\textbf{Criterion:} The population is different on study-relevant criteria

\textbf{Example:} Using domain experts with industry experience rather than university students where expertise is crucial &
\textbf{Criterion:} The approach to data analysis is different

\textbf{Example:} Video data may be analyzed for quantitative metrics rather than qualitative themes \\
\bottomrule

\end{tabularx}
\caption{Three comparison levels across four dimensions with criteria and examples.}
\label{tab:component-comparison-examples}
\end{table*}

\subsection{Dimension Space}
To enable comparison and analysis of replication studies across a range of empirical domains, the REPVIS framework describes the relationship between a new study and a reference study through four practical design dimensions: \colorbox[HTML]{B3E2CD}{Experiment}, \colorbox[HTML]{FDCDAC}{Data}, \colorbox[HTML]{CBD5E8}{Participant}, and \colorbox[HTML]{F4CAE4}{Analysis}.

We choose these dimensions because together they represent the practical-level study cycle~\cite{cutler2026revisit, purchase2012experimental} and cover the evidence chain decisions made for the replication study: \colorbox[HTML]{B3E2CD}{Experiment} specifies the process of gathering \textbf{evidence}; \colorbox[HTML]{FDCDAC}{Data} specifies the nature of the \textbf{evidence}; \colorbox[HTML]{CBD5E8}{Participant} specifies the population and sampling characteristics that determine who provides the \textbf{evidence}; and \colorbox[HTML]{F4CAE4}{Analysis} specifies the inferential procedures through which \textbf{evidence} is transformed into claims of the study.

\noindent\colorbox[HTML]{B3E2CD}{Experiment} describes the methodology, protocol and procedure used to gather evidence, such as tasks, conditions and stimuli.


\noindent\colorbox[HTML]{FDCDAC}{Data} refers to the evidence source on which the study claims are ultimately based, such as answers, logs and response time.

\noindent\colorbox[HTML]{CBD5E8}{Participant} is defined as the population and sampling specifications of a study, including factors such as recruitment source, inclusion/exclusion criteria and expertise/experience requirements.

\noindent\colorbox[HTML]{F4CAE4}{Analysis} method defines the procedures used for transforming the evidence collected into claims. 

In practice, other factors may also shape design, such as the research team and research question. While the research team conducting the replication study may be the same as, or different to, the team who did the reference study, this fact is only interesting if there is suspected bias. With respect to the research question, it is assumed that the reference and replication studies are addressing the same (or similar) research questions -- since if this were not the case, the latter would simply be a new and different study.

While the \emph{motivation} for a replication study is not part of its \emph{design}, it is relevant when considering the research question, and how it may affect design. If the purpose is for \emph{validation} (that is, simply confirming that previously published results are indeed correct), the expectation is that all dimensions -- as much as possible -- will be unchanged, as would be the case if validating the reaction from mixing two chemicals. If the purpose is for \emph{generalization} (for example, seeing whether previous results hold for different contexts), then clear decisions will be made regarding how dimensions will differ. Our framework can guide this decision-making.

\subsection{Comparison Level}
The dimension space addresses \emph{what} is compared; the comparison level specifies \emph{how} comparisons are made. For each dimension pair (new vs. reference), we use three comparison levels (\emph{identical}, \emph{similar}, and \emph{different}) to capture structural comparability and correspondence (Table~\ref{tab:component-comparison-examples}). These levels are descriptive rather than normative; they do not imply one choice is "better" than another.

\noindent\textbf{\textit{Identical}} The dimension serves the same function and produces the same type of evidence as in the reference study. Unavoidable implementation differences may exist, but these are unintentional. Comparison is \textbf{direct} and like-for-like without extra interpretation.

\noindent\textbf{\textit{Similar}} The dimension differs substantively, so \textbf{direct} comparison is inappropriate. However, the dimension still plays a comparable function and yields a comparable type of evidence. So \textbf{comparability} is preserved. 

\noindent\textbf{\textit{Different}} The dimension is so different that neither \textbf{direct} nor meaningful \textbf{comparability} is preserved, although the dimension is still important to the experiment design.

We use three comparison levels rather than a binary or a more fine-grained metric. A binary "same/different" split cannot capture "different-yet-comparable" cases (e.g., cross-validation vs. triangulation). Conversely, finer-grained schemes may increase descriptive precision, but will result in labeling inconsistencies and may make patterns difficult to identify.


\subsection{Framework and Collapse rules}
\label{sec:collapse_rules}

\begin{figure}[htb]
  \centering
  \includegraphics[width=1.0\linewidth]{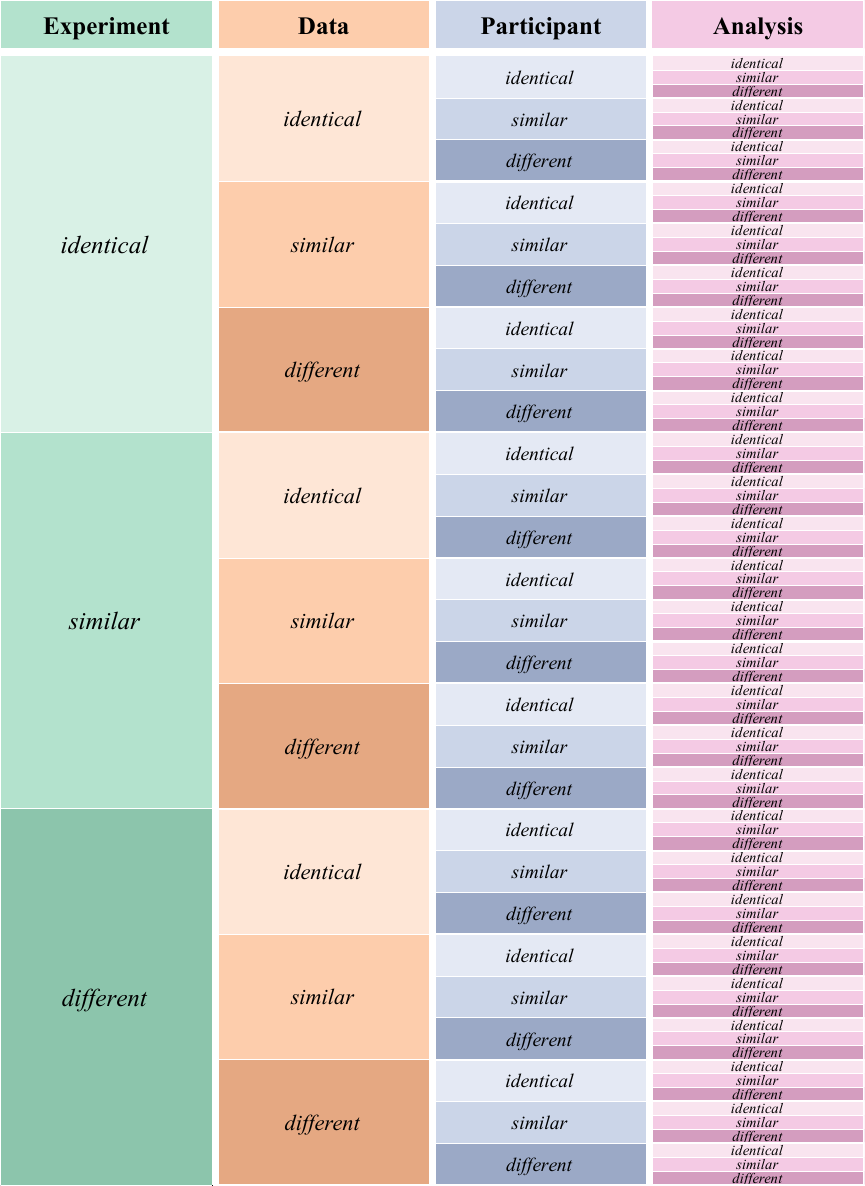}
  \caption{\label{fig:firstExample}
          The REPVIS replication design space, enumerating all 81 possible replication designs. The column order holds no prescriptive meaning, since real designs are often iterative and intertwined across dimensions. "Collapse rules" in \S\ref{sec:collapse_rules} describe where columns may be discarded and ignored in specific circumstances.}
\end{figure}

REPVIS instantiates the three comparison levels across the four dimensions and organizes their combinations into a framework (Figure \ref{fig:firstExample}), where 81 design possibilities are enumerated: the table makes the combinatorial design space explicit. Each of these 81 possibilities encodes a point in multi-dimensional design space that describes how a replication study relates to a reference study across dimensions:
\[
(\colorbox[HTML]{B3E2CD}{$E_r$} \times \colorbox[HTML]{FDCDAC}{$D_r$} \times \colorbox[HTML]{CBD5E8}{$P_r$} \times \colorbox[HTML]{F4CAE4}{$A_r$})
\quad r \in \{\textit{identical},\textit{similar},\textit{different}\}.
\]

As in many design spaces, enumerating all possibilities reveals some impossible designs, which we define by "collapse rules".

\noindent\textbf{\colorbox[HTML]{FDCDAC}{$D_{identical}$} is used as evidence}. If the new study reuses the exact data from the reference study as its \colorbox[HTML]{FDCDAC}{D}, then clearly the original experimental method and participants were used to create this data -- that is, no design choice about the experimental process or nature of the participants can be made (the \colorbox[HTML]{B3E2CD}{E} and \colorbox[HTML]{CBD5E8}{P} columns are collapsed); the only option is to choose the nature of \colorbox[HTML]{F4CAE4}{A}.

\noindent\textbf{Computational}. The flexibility of REPVIS also encompasses computational study designs  (e.g., benchmark, simulation, or algorithm). In this case, no participants are involved and so the \colorbox[HTML]{CBD5E8}{P} column is collapsed.

\subsection{Worked Examples}
\label{sec:worked_examples}
We provide four worked examples to show the two-fold purpose of REPVIS: first, to support clear retrospective characterization of existing replication studies (examples 1 \& 2); second, to support the planning process for new future replication studies (examples 3 \& 4). All examples are replications of the classic Cleveland \& McGill study~\cite{cleveland1984graphical} which asked participants to judge scale and length in different presentations of bar and pie charts. 

\noindent\textbf{Example 1}: $\colorbox[HTML]{B3E2CD}{$E_{\textit{similar}}$}\times\colorbox[HTML]{FDCDAC}{$D_{\textit{similar}}$}\times\colorbox[HTML]{CBD5E8}{$P_{\textit{different}}$}\times\colorbox[HTML]{F4CAE4}{$A_{\textit{identical}}$}$
\newline
Khalaila et al.~\cite{khalaila2025they} evaluated graphical perception of tactile graphics using swell-form printing. The core experimental paradigm is similar, but the population shifts to blind or low-vision (BLV) and the paper provides guidelines for BLV visualization design. This study shows the potential for re-assessing prior findings for different user groups.

\noindent\textbf{Example 2}: $\colorbox[HTML]{B3E2CD}{$E_{\textit{similar}}$}\times\colorbox[HTML]{FDCDAC}{$D_{\textit{similar}}$}\times\colorbox[HTML]{CBD5E8}{$P_{\textit{similar}}$}\times\colorbox[HTML]{F4CAE4}{$A_{\textit{different}}$}$
\newline
Davis et al.~\cite{davis2022risks} revisited the comparison task with a new study, keeping a comparable protocol and a broadly similar participant pool. The key change is the analysis method \colorbox[HTML]{F4CAE4}{$A$}: instead of an average-observer summary, they use Bayesian multilevel modeling to foreground individual differences and reinterpret the canonical ranking. This study demonstrates how different \colorbox[HTML]{F4CAE4}{$A$} can be used to propose new explanations and extend prior results.

\noindent\textbf{Example 3}: $\colorbox[HTML]{B3E2CD}{$E_{\textit{identical}}$}\times\colorbox[HTML]{FDCDAC}{$D_{\textit{similar}}$}\times\colorbox[HTML]{CBD5E8}{$P_{\textit{identical}}$}\times\colorbox[HTML]{F4CAE4}{$A_{\textit{identical}}$}$
\newline
A new study might conduct a precise replication of the graphical perception study by matching the protocol and analysis, with only unavoidable implementation differences (for example, physical location, individual participants). The goal is an unambiguous validation of the original findings. Note that the data is not identical (since that would mean using the exact same data as the original study); it is similar since it is in the same form as the original study.

\noindent\textbf{Example 4}: $\colorbox[HTML]{B3E2CD}{$E_{\textit{different}}$}\times\colorbox[HTML]{FDCDAC}{$D_{\textit{different}}$}\times\colorbox[HTML]{CBD5E8}{$P_{\textit{similar}}$}\times\colorbox[HTML]{F4CAE4}{$A_{\textit{similar}}$}$
\newline
A new study could triangulate the graphical perception study using a psychophysics threshold paradigm (e.g., 2AFC with an adaptive staircase) instead of magnitude estimation. This makes the experimental design \colorbox[HTML]{B3E2CD}{$E$} and the nature of the data \colorbox[HTML]{FDCDAC}{D} different, but other dimensions remain comparable. The analysis can still remain comparable by mapping thresholds (or derived sensitivity measures such as $d'$) to an accuracy-style interpretation, yielding triangulation for original findings through an alternative study paradigm.

\section{Discussion}

Our REPVIS framework is the first attempt to identify and delineate the decisions required for conducting a replication study. Prior work has taken three approaches:

\noindent\textbf{Category 1: advocacy, terminology and high-level taxonomy.}
Conceptual categorizations and terminology-level discussions~\cite{kosara2018skipping,quadri2019you,hornbaek2014once,acm_artifact_review_badging_2020} are useful for framing the landscape, but provide limited actionable and consistent descriptions of practical changes and comparisons. Many of these notions can be expressed within REPVIS. For example, Kosara's \emph{reanalysis}~\cite{kosara2018skipping} can be written as $\colorbox[HTML]{FDCDAC}{$D_{\textit{identical}}$}\times\colorbox[HTML]{F4CAE4}{$A_{\textit{identical/similar/different}}$}$; \emph{direct replication} as $\colorbox[HTML]{B3E2CD}{$E_{\textit{identical}}$}\times\colorbox[HTML]{FDCDAC}{$D_{\textit{similar}}$}\times\colorbox[HTML]{CBD5E8}{$P_{\textit{identical}}$}\times\colorbox[HTML]{F4CAE4}{$A_{\textit{identical}}$}$; and \emph{conceptual replication} as $\colorbox[HTML]{B3E2CD}{$E_{\textit{similar/different}}$}\times\colorbox[HTML]{FDCDAC}{$D_{\textit{similar/different}}$}\times\colorbox[HTML]{CBD5E8}{$P_{\textit{identical/similar/different}}$}\times\colorbox[HTML]{F4CAE4}{$A_{\textit{identical/similar/different}}$}$.

\noindent\textbf{Category 2: practice-oriented guidelines.}
Practical guidance, such as that provided by Sukumar and Metoyer~\cite{sukumar2018towards} extracts recommendations from published replication practices, but largely takes the form of experience-based checklists with no unified structural representation to facilitate description or cross-study comparison -- this structure is provided by the REPVIS.

\noindent\textbf{Category 3: design space views from other fields.}
In other scientific domains, for example, work by Patil et al.~\cite{patil2016statistical} decomposes the scientific process into dimensions (e.g., experiment, data, analyst, analysis method) and use a binary "same/different" comparison levels. While valuable for concept definition, the coarse granularity of this approach does not capture the common case of being "different-yet-comparable" ("similar" in REPVIS), which is particularly important for replication studies whose motivation is generalization. In addition, other scientific fields may lack the complexity of recruiting participants \colorbox[HTML]{CBD5E8}{P}.

Our REPVIS framework is a concrete first step towards being able to both characterize the landscape of existing replication studies, as well as providing guidance to researchers wishing to investigate the validity or generalisability of published results. We expect it to evolve with use in either endeavour. Future work issues to consider are the nature of (and motivation for) the research question, studies that are extensions of prior work (not simple and pure replications), distinguishing sub-dimensions of the dimension space (in particular, the \colorbox[HTML]{B3E2CD}{Experiment} dimension), and considering the appropriate granularity and combination of the three comparison levels. While our initial framework has been proposed on a conceptual basis, its evolution will be data-driven, based on evaluating the reality of actual replication research.

\section{Conclusion}
We have defined an exhaustive design space (REPVIS) for experimental replication research studies which considers the extent to which the experiment design, the dependent data, the participant demographics and the data analysis methods of the new (replication) study align with the original (reference) study. Our contribution of a novel framework enables unambiguous comparison between two experiments which address the same or similar research questions, and, in doing so, removes the need for confusing and uncertain terminology. As a multi-dimensional space, the framework is a useful tool for researchers wishing to revisit a prior study for the purposes of replicating its results in various different contexts and ways, since it enumerates the wide range of design possibilities.

\bibliographystyle{eg-alpha-doi} 
\bibliography{reference}       



\end{document}